\documentclass[pra,aps,showpacs,twocolumn,groupedaddress,superscriptaddress,floatfix]{revtex4}
\usepackage{epsfig}
\usepackage{graphicx}
\usepackage{amsfonts}
\usepackage{amsmath}

\begin{document}
\title{Two-photon interference with two independent pseudo-thermal sources }
\author{Yan-Hua Zhai$^1$, Xi-Hao Chen$^{1, 2}$ and Ling-An Wu}
\thanks{Corresponding author: wula@aphy.iphy.ac.cn}
\affiliation{Laboratory of Optical Physics, Institute of Physics,
Chinese Academy of Sciences, Beijing 100080,
China\\\it{$^2$}Department of Physics, Liaoning University, Shenyang
110036, China}

\begin{abstract}
The nature of two-photon interference is a subject that has aroused
renewed interest in recent years and is still under debate. In this
paper we report the first observation of two-photon interference
with independent pseudo-thermal sources in which sub-wavelength
interference is observed. The phenomenon may be described in terms
of optical transfer functions and the classical statistical
distribution of the two sources.
\end{abstract}
\pacs{42.50.Ar, 42.50.Dv, 42.25.Hz, 42.50.St} \maketitle

Young's double-slit experiment is one of the most important
experiments in the history of physics,  being the earliest
demonstration of the interference of wave motion. Later, it also
provided powerful evidence for the wave-particle duality of light. A
phenomenon of profound significance, interference is also ubiquitous
in all areas of physics, but the term is by default understood to
mean first-order interference, i.e. interference observed in the
first-order intensity of the field in question. It was only after
more than 150 years that effects due to second-order intensity
correlations in optics was first considered and made use of by
Hanbury Brown and Twiss in their landmark experiment~\cite{hbt} to
measure the angular diameters of stars with an accuracy far
surpassing that achievable by the Michelson interferometer because
of its insensitivity to phase disturbances.

At the end of the last century the successful demonstration of
two-photon interference with entangled light produced by spontaneous
parametric down-conversion (SPDC)~\cite{prl82, prl85, pra64, prl87,
prl90,RMP71, Rpp66} brought to attention the question of whether
two-photon interference can be considered as the interference of two
distinct photons~\cite{prl77, pra57, quant07, quant08}. Recently,
there has been great interest in two-photon interference with
thermal light~\cite{pra70, ol29, pra70l,epl68, jun, prl94, pra72,
prl96} but the nature of two-photon interference is still under
debate and so deserves further research.

After Mandel \emph{et al.} performed their famous classical
first-order correlation interference experiment with two independent
lasers~\cite{nature, pr159} in the sixties, the question of
interference between independent beams became widely
discussed~\cite{PR134, pr172, PRD1, rmp58}. In Ref.~\cite{ajp32}, a
classical-like first-order interference-diffraction pattern was
obtained with two independent pseudo-thermal light sources.
Recently, nonclassical two-photon interference effects were observed
with one photon coming from SPDC and the other from a weak laser
source~\cite{ptrsla, prl2003}, while Kaltenbaek \emph{et al.}
succeeded in observing interference of independent photons produced
by two SPDC sources~\cite{quant}. However, sub-wavelength
second-order interference was not reported. In this paper we
describe the first observation of two-photon interference with two
independent pseudo-thermal point sources which exhibited
sub-wavelength interference.

\begin{figure}
\includegraphics[width=10cm]{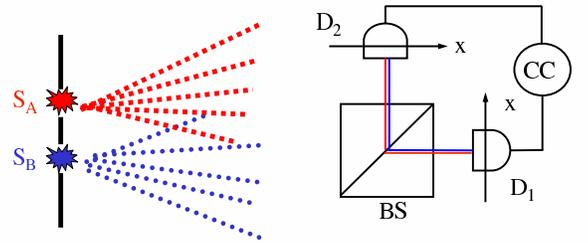}
\caption{Principle of the experiment. $\mathrm{S_A}$,
$\mathrm{S_B}$: two independent pseudo-thermal sources; BS:
non-polarizing beam splitter; $\mathrm{D_1}$, $ \mathrm{D_2}$:
single photon detectors; CC: coincidence counter; $x$: scan
direction of detectors} \label{sketch}
\end{figure}

The principle of the experiment is shown in Fig.~\ref{sketch}. Two
independent pseudo-thermal light sources $\mathrm{S_A}$ and
$\mathrm{S_B}$ are located at two pinholes. The beams from the two
pinholes pass through a beam splitter and are detected by two
single-photon detectors $\mathrm{D_1}$ and $\mathrm{D_2}$,
respectively, which can be translated in the $x$ directions. The
output signals are sent to a coincidence counter. The experiment is
first  performed with both sources having the same polarization and
then with perpendicular polarizations.

\begin{figure}
\includegraphics[width=10cm]{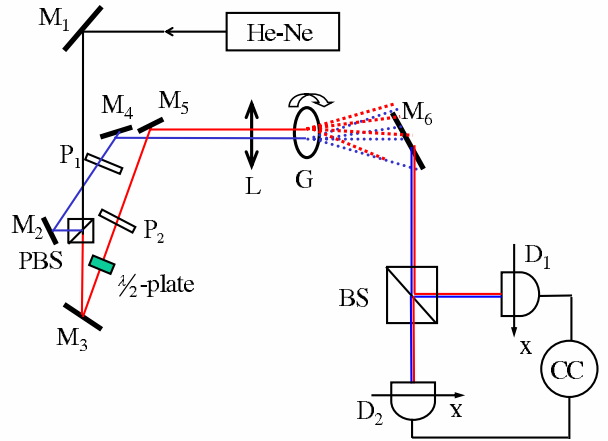} \caption{Experimental set-up of interference with two
independent light sources of the same polarization. $\mathrm{M_1}$ -
$\mathrm{M_6}$: mirrors; PBS: polarizing beam splitter; BS:
non-polarizing beam splitter; $\mathrm{P_1}$, $ \mathrm{P_2}$:
polarizers; L: lens (f=10mm); G: rotating ground glass plate
(12Hz).} \label{set-up}
\end{figure}

An outline of the experimental set-up is shown in Fig.~\ref{set-up}.
The stabilized He-Ne laser of wavelength 632.8nm and length
approximately 20cm (Model FS100, Beijing Fangshi Keji Co) produces
two longitudinal modes of perpendicular polarization, with a
frequency difference of 1 GHz. It has been shown that two such
adjacent perpendicular modes have no phase correlation and so are
independent of each other~\cite{note, ajp32}. They are separated by
a $50/50\%$ polarizing beam splitter ($\mathrm{PBS}$) so that one
mode is reflected and the other transmitted by the PBS. The
reflected beam passes from mirror $\mathrm{M_2}$ through polarizer
$\mathrm{P_1}$ to mirror $\mathrm{M_4}$, and is $s$-polarized. The
$p$-polarized transmitted beam is reflected by mirror
$\mathrm{M_3}$, then passes through a half-wave plate and polarizer
$ \mathrm{P_2}$ before being reflected by mirror $\mathrm{M_5}$ to
emerge parallel to the other beam. It may be converted to
$s$-polarization by rotating the half-wave plate and polarizer
$\mathrm{P_2}$.  The two beams are then focused by lens L at two
spots A and B separated by about 1.1mm on a ground glass plate which
rotates at a speed of 12Hz. The diameter of the spots is about
0.11mm, so they are equivalent to two pseudo-thermal pinhole light
sources. Light scattered from the two spots is then reflected by
mirror $ \mathrm{M_6}$ and divided by a $50/50\%$ non-polarizing
beam splitter BS. The reflected and transmitted beams are detected
by single-photon detectors $ \mathrm{D_1}$ and $\mathrm{D_2}$
(Perkin Elmer SPCM-AQR-13), respectively. The output pulses from the
two detectors are sent to a coincidence counting circuit.

\begin{figure}
\includegraphics[width=5cm]{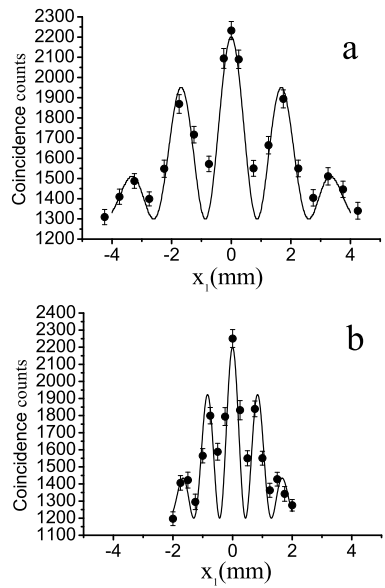} \caption{Coincidence counts in 0.1 second. (a) As a
function of the position of $\mathrm{D_1}$ with $\mathrm{D_2}$
fixed. (b)As a function of the position of detectors $\mathrm{ D_1,
D_2}$ when they were scanned in opposite directions ($x, -x$)
simultaneously. The solid curves are theoretical plots.}
\label{sub-wavelength}
\end{figure}

To begin with, detector $\mathrm{D_2}$ was kept fixed while $
\mathrm{D_1}$ was scanned in the horizontal direction and the rate
of coincidence counts recorded as a function of its position. As can
be seen from Fig.~\ref{sub-wavelength}a, a classical-like
first-order interference-diffraction pattern can be obtained. The
distance between the zeroth-order and the first-order interference
peak is about 1.7mm.

Next, when the detectors $\mathrm{D_1}$ and $\mathrm{D_2}$ were
scanned in \emph{opposite} directions $(x, -x)$ in steps of 0.25mm
simultaneously, the second-order interference-diffraction pattern
shown in Fig.~\ref{sub-wavelength}b was obtained. The distance
between the zeroth-order and the first-order interference peak is
about 0.85mm, which is exactly half that of the classical case. This
is very similar to the sub-wavelength effect, which was first
predicted and observed for two-photon interference with entangled
photon pairs~\cite{prl85, prl87}, then was recently observed with a
pseudo-thermal source~\cite{epl68, jun} as well as with a single
true thermal source~\cite{pra72}.

The coincidence count rate is proportional to the second-order
correlation function
\begin{multline}
\label{G2}
G^{(2)}(x_1, t_1, x_2, t_2)\\
\shoveleft = \langle \hat{E}_{2}(x_2, t_2)^{(-)} \hat{E}_{1}(x_1,
t_1)^{(-)} \hat{E}_{1}(x_1, t_1)^{(+)} \hat{E}_{2}(x_2, t_2)^{(+)}
\rangle,
\end{multline}
where $|\psi(t)\rangle$ is the state of the system, and
$\hat{E}_{i}(x_i, t_i)^{(+)}$, $ \hat{E}_{i}(x_i, t_i)^{(-)}$ are
the positive and negative frequency field operators at time $t_i$ at
detectors $\mathrm{D}_i (i=1,2)$ located at $x_i$, respectively.

We will now derive a simple explanation for this sub- wavelength
interference with two independent sources. The transmission function
of source A of the double-source function can be written as
\begin{equation}
T_A(x_0)=\begin{cases} 1, & (d-s)/2\leq x_0 \leq (d+s)/2, \\
0, &\text{otherwise}
\end{cases}
\end{equation}
and for source B,
\begin{equation}
T_B(x_0)=\begin{cases} 1, & -(d+s)/2\leq x_0 \leq -(d-s)/2, \\ 0,
&\text{otherwise}
\end{cases}
\end{equation}
where $d$ is the distance between the two spots,  $s$ is their
diameter, and $x_0$ is the distance from the central point between
them.

After the beam from source A is divided at BS and detected at
$\mathrm{D_1}$ and $\mathrm{D_2}$, the registered coincidence count
is proportional to the second-order correlation function, and for
Gaussian thermal fields, the relationship between the second- and
first- order correlation functions $G^{(2)}$ and $G^{(1)}$ is given
by~\cite{goodman}
\begin{eqnarray}
\label{G(2)_AB} \lefteqn{G^{(2)} (x_1,
x_2)}\nonumber\\
\hspace{-5cm}&=&\langle\hat{b}^\dag(x_1)\hat{b}^\dag(x_2)
\hat{b}(x_2)\hat{b}(x_1)\rangle\nonumber\\
&=&|\langle \hat{b}^\dag(x_1)\hat{b}(
x_2)\rangle|^2\nonumber\\
& &+\langle\hat{b}^\dag(x_1)\hat{b}(x_1) \rangle\langle
\hat{b}^\dag(x_2)\hat{b}(x_2)\rangle\nonumber
\\&=&|G^{(1)}(x_1,
x_2)|^2+\langle G^{(1)}(x_1,x_1)G^{(1)}(x_2,x_2)\rangle
\end{eqnarray}
where $ \hat{b}^\dag(x_i)$ and $\hat{b}(x_i)$ are the creation and
annihilation operators at detectors $\mathrm{D}_i$ located at
$(x_i)$, respectively.

From the Wiener-Khintchine theorem~\cite{mandel}, the first-order
spectral correlation  for thermal light satisfies
\begin{equation}
\label{spectrum} \langle \hat{a}^\dag(q_1)\hat{a}(q_2)\rangle=S(q_1)
\delta(q_1- q_2)
\end{equation}
where $S(q_1)$ is the spatial spectral distribution and $q$ is the
transverse wave vector of the optical field. The spectral width of
thermal light can be assumed to be infinite, so $S(q_1)=1$.

For source A, we can calculate the first-order correlation function
by using equation~(\ref{spectrum})
\begin{eqnarray}
\label{G(1)} \lefteqn{G^{(1)}_A(x_1, x_2)}\nonumber\\
\hspace{-5cm}&=&\langle\hat{b}^\dag_A(x_1)\hat{b}_A(x_2)\rangle\nonumber\\&=&
\int\int\tilde{H}^*_A(x_1,-q_1)\tilde{H}_A(x_2,-q_2)\langle\hat{a}^\dag_A(q_1)\hat{a}_A(q_2)
\rangle dqdq'\nonumber\\&=&\int
\tilde{H}^*_A(x_1,-q_1)\tilde{H}_A(x_2,-q_1)dq   .
\end{eqnarray}
Here $ \hat{a}^\dag_A(q_1)$ and $\hat{a}_A(q_2)$ are the creation
and annihilation operators for the source A, and
$\tilde{H}_A(x_i,-q_i)$ is the partial Fourier transform of the
impulse response function from the pseudo-thermal source A to the
detectors $\mathrm{D}_i$:
\begin{eqnarray}\label{fu}
\lefteqn{\tilde{H}_A(x_i, -q_i)}\\
\hspace{-5cm}&=&\frac{1}{\sqrt{2\pi}}\int \int h_f(x_i, x')h_{A}(x',
x_0)dx'\mathrm{exp}[iq_ix_0]dx_0\nonumber,
\end{eqnarray}
wwhere $h_{A}(x', x_0)=T_A(x_0)\delta (x'-x_0)$ is the impulse
response function for the upper spot of the double-source and $
h_f(x_i, x')$ is the impulse response function in free space from
the source to the detectors $\mathrm{D}_i$. Substituting
equations~(\ref{spectrum}) and~(\ref{fu}) into
equation~(\ref{G(1)}), we can obtain
\begin{eqnarray}
\label{G(1)_A}
\lefteqn{G^{(1)}_A(x_1, x_2)}\nonumber\\
\hspace{-5cm}&=&\frac{k}{2\pi z}\frac{1}{2\pi}\int\int\int
T_A(x'_0)T_A(x_0)\delta (x'-x_0)\nonumber\\ & &
\mathrm{exp}[i(\frac{kx_1}{z}-q)x'_0-i(\frac{kx_2}{z}-q)x_0dqdx_0dx'_0\nonumber\\&=&\frac{k}{2\pi
z}\int^{\frac{d+s}{2}}_{\frac{d-s}{2}}\mathrm{exp}[i\frac{k}{z}(x_1-x_2)x_0]dx_0\nonumber\\
&=&\frac{1}{\pi(x_1-x_2)}\{\mathrm{cos}[\frac{k(x_1-x_2)d}{2z}]\mathrm{sin}[\frac{k(x_1-x_2)s}{2z}]\nonumber\\&&
-i\mathrm{sin}[\frac{k(x_1-x_2)d}{2z}]\mathrm{sin}[\frac{k(x_1-x_2)s}{2z}]\}
\end{eqnarray}
where $z$ is the distance to the detector and $\lambda$ is the
wavelength of the pseudo-thermal light.

For the lower pseudo-thermal source B we obtain a similar expression
but with a plus instead of a minus sign before the second term.
\begin{eqnarray}
\label{G(1)_B} \lefteqn{G^{(1)}_B(x_1, x_2)}\nonumber\\
\hspace{-5cm}&=&\frac{1}{\pi(x_1-x_2)}\{\mathrm{cos}[\frac{k(x_1-
x_2)d}{2z}]\mathrm{sin}[\frac{k(x_1-x_2)s}{2z}]\nonumber\\&& +
i\mathrm{sin}[\frac{k(x_1-x_2)d}{2z}]\mathrm{sin}[\frac{k(x_1-x_2)s}{2z}]\}.
\end{eqnarray}

If both sources A and B have the same polarization, we can calculate
the second-order correlation function from equation ~(\ref{G(2)_AB})
to be
\begin{eqnarray}
\label{same}
\lefteqn{G^{(2)}(x_1, x_2)}\nonumber\\
&=&
\langle[\hat{b}^\dag_A(x_1)+\hat{b}^\dag_B(x_1)][\hat{b}^\dag_A(x_2)+
\hat{b}^\dag_B(x_2)]\nonumber\\&&
[\hat{b}_A(x_2)+\hat{b}_B(x_2)][\hat{b}_A(x_1)
+\hat{b}_B(x_1)]\rangle\nonumber
\\&=&|G^{(1)}_A(x_1,x_2)+G^{(1)}_B(x_1,x_2)|^2+
G^{(1)}_A(x_1,x_1)G^{(1)}_A(x_2,x_2)\nonumber\\&&
+G^{(1)}_A(x_1,x_1)G^{(1)}_B(
x_2,x_2)+G^{(1)}_B(x_1,x_1)G^{(1)}_A(x_2,x_2)\nonumber\\&&
+G^{(1)}_B(x_1,x_1) G^{(1)}_B(x_2,x_2).
\end{eqnarray}
Here $\hat{b}^\dag_m(x_i)$ and $\hat{b}_m( x_i)$ are the creation
and annihilation operators for the source $m (m=A,B)$ at detectors
$\mathrm{D}_i$ located at $(x_i)$, respectively. By using
equations~(\ref{G(1)_A}) and~(\ref{G(1)_B}), the second- order
correlation function when the detectors are scanned in opposite
directions $(x, -x)$ can thus be written as
\begin{eqnarray}
\label{resultsame} G^{(2)}(x,-x)=(\frac{ks}{\pi
z})^2[1+\mathrm{sinc}^2\frac{\pi sx}{(\lambda/2)z}
\mathrm{cos}^2\frac{\pi dx}{(\lambda/2)z}] .
\end{eqnarray}
The sub-wavelength interference pattern thus originates from the
$(\lambda/2)$ term in equation~(\ref{resultsame}).

However, when the beams from the two pseudo-thermal sources are
perpendicularly polarized to each other, the second-order
correlation function from equation ~(\ref{G(2)_AB}) is
\begin{eqnarray}
\label{different} \lefteqn{G^{(2)}(x_1, x_2)}\nonumber\\&=&
\langle\hat{b}^\dag_A(x_1)\hat{b}^\dag_A(x_2)\hat{b}_A(x_2)\hat{b}_A(x_1)
\rangle\nonumber\\&&+\langle\hat{b}^\dag_B(x_1)\hat{b}^\dag_B(x_2)\hat{b}_B(
x_2)\hat{b}_B(x_1)\rangle\nonumber\\&&+\langle\hat{b}^\dag_A(x_1)\hat{b}^\dag_B(x_2)\hat{b}_B(x_2)\hat{b}_A(x_1)
\rangle\nonumber\\&&+\langle\hat{b}^\dag_B(x_1)\hat{b}^\dag_A(x_2)\hat{b}_A(
x_2)\hat{b}_B(x_1)\rangle\nonumber\\
&=&|G^{(1)}_A(x_1,x_2)|^2+G^{(1)}_A(x_1,x_1)G^{(1)}_A(x_2,x_2)\nonumber\\&&
+|G^{(1)}_B(x_1,x_2)|^2+G^{(1)}_B(x_1,x_1)
G^{(1)}_B(x_2,x_2)\nonumber\\&&+G^{(1)}_A(x_1,x_1)
G^{(1)}_B(x_2,x_2)\nonumber\\&&+G^{(1)}_B(x_1,x_1)
G^{(1)}_A(x_2,x_2).
\end{eqnarray}
Substituting equations~(\ref{G(1)_A}) and~(\ref{G(1)_B}) into the
above equation we obtain
\begin{eqnarray}
\label{result} G^{(2)}(x,-x)=(\frac{ks}{\pi z})^2[1+\frac{1}{2}
\mathrm{sinc}^2\frac{\pi sx}{(\lambda/2) z} ] .
\end{eqnarray}
For this situation we can see that the second-order correlation
function is only a superposition of the Hanbury Brown and Twiss
effect~\cite{hbt} from each pseudo-thermal source, and no
interference pattern is observable. This was confirmed in the
following experiment.

\begin{figure}
\includegraphics[width=5cm]{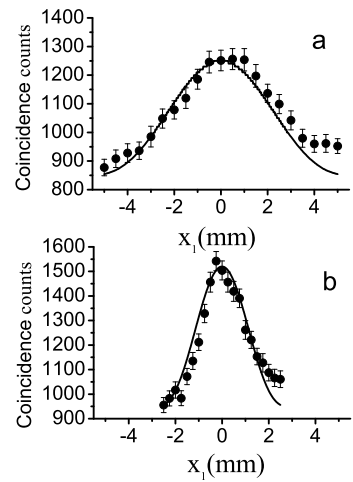}
\caption{Coincidence counts in 0.1 s for perpendicularly polarized
beams: (a) As a function of the position of $\mathrm{D_1}$ with
$\mathrm{D_2}$ fixed; (b) As a function of the position of detectors
$\mathrm{ D_1, D_2}$ when they were scanned in opposite directions
($x, -x$) simultaneously. The solid curves are theoretical plots. }
\label{perpendicular}
\end{figure}

We removed the half-wave plate and adjusted the polarizer $P_2$ in
Fig.~\ref{set-up} so that the two sources A and B emitted light of
orthogonal polarizations. With $\mathrm{D_2}$ fixed, we scanned
$\mathrm{D_1}$ and obtained the results shown in
Fig.~\ref{perpendicular}a, while Fig.~\ref{perpendicular}b shows the
plot obtained when the two detectors were scanned in \emph{opposite}
directions $(x, -x)$ simultaneously. The solid curves are
theoretical plots calculated from equation~(\ref{result}). It is
evident that there is no interference-diffraction pattern in either
case.

This result would seem to be obvious from the point of view of
classical first-order interference-diffraction. On the other hand,
from the quantum aspect when the two sources have orthogonal
polarizations which photons come from which source can be
distinguished and so no interference is possible. However, when the
two sources have the same polarization we cannot distinguish which
photons come from which source and so interference is observed. Our
experimental results have thus confirmed that indistinguishability
is the reason behind interference, even in the case of two sources
that lack coherence in the usual sense.

Over the last few decades our understanding of interference, one of
the most important concepts of physics, has advanced considerably
since the days that Dirac said~\cite{dirac},``Each photon then
interferes only with itself. Interference between two different
photons can never occur." The statement provoked widespread debate
and led to a surge of experimental tests as well as philosophical
argument. It is now generally agreed that Dirac's statement should
be viewed in its historical content when the resolution time of
photon detectors was still limited. As a way to resolve the
misunderstanding, Shih \emph{et al.} indicate that ``two-photon
correlation interference is the result of each pair of independent
photons interfering with itself" ~\cite{epl68}, and they also
maintain that two-photon interference with thermal light is not
caused by the statistical correlation of the intensity
fluctuations~\cite{prl96}, even if the results can be obtained by a
classical or quantum derivation. However, there is still no
consensus on the actual mechanism behind this phenomena. In
first-order interference it is the phase difference in the field
amplitudes, caused by the different path lengths to the point of
detection, that is the origin of the interference. Similarly, it is
the phase difference of the two-photon amplitudes due to different
paths to the two points of detection that gives rise to second-order
correlation interference. Nonetheless, regardless of whether we are
able to observe it or not, interference is an ever-present
phenomenon of nature.

In summary, we have observed sub-wavelength interference with two
independent pseudo-thermal sources, which may be helpful for
understanding the nature of two- photon interference. There is still
much to be explored regarding the properties of thermal light
sources although they have been around for a long time. It is even
conceivable that thermal light may find special applications in
optical imaging and other fields because of its two-photon
correlation characteristics~\cite{prl96, prl92}.

We thank D. Zhang for helpful discussions. This work was supported
by the Natural Science Foundation of China Grant No. 60578029, and
the National Program for Basic Research in China Grant No.
001CB309301.

\end{document}